# THE INFLUENCE OF ENVIRONMENTAL EFFECTS ON GALAXY FORMATION

Richard G. Bower*
*Royal Observatory, Blackford Hill, Edinburgh, EH9 3HJ, U.K.*

**ABSTRACT.** These notes collect together current work on the effect of the environment on galaxy formation and evolution. They are broken into four distinct parts. The first deals with the observational debate surrounding the question of whether galaxies are initially created the same and have only recently been modified by their environment, or whether galaxy formation itself is strongly perturbed by the local neighbourhood. The second section deals with observations of distant galaxy clusters. These provide direct evidence that some galaxies, at least, have changed substantially over the last 5 Gyr. The third and fourth parts look at theoretical work on this theme. They study models for the origin of the density–morphology relation, and the impact of environmentally modulated galaxy formation on the large-scale structure of the universe.

## Introduction

The existence of a well-defined, perhaps even universal, density–morphology relation (Hubble & Humason, 1931, Dressler, 1980) demonstrates that galaxy evolution is strongly influenced by the environment. Since I have been careful to distinguish between galaxy evolution and formation, some people will disagree. It is more likely, they would argue, that galaxies differ between environments because of the way in which they formed. This is a fundamental question, and I believe one which we are now close to answering: are all galaxies initially formed by the same proceeses, only recently having evolved to take on the observed variety of morphologies; or does their initial environment so affect their formation that their intrinsic differences are imprinted from the beginning? In these notes I have attempted to sketch the recent research surrounding this question, and also to draw in the wider issue of the link between 'biased' galaxy formation (eg., Kaiser, 1984) and the large-scale structure of the universe. From the start, it seems seems obvious that *bias* (ie., the differential way in which luminous matter traces the underlying gravitational structure of the universe) cannot be understood without first solving the mysteries of the density–morphology relation.

I have split these notes into 4 parts. The first draws together information from detailed studies of nearby galaxies. The discussion here is essentially observational, as I present the evidence for and against the 'nature' and 'nurture' hypotheses for the origin of early-type galaxies. Because of my own background in this field, the discussion concentrates on the stellar populations of E and S0 galaxies. I have included relatively little discussion of spiral galaxies or of the dynamical structure of ellipticals. These are important points that I hope will be amply discussed during the meeting.

---

* *Present address: Dept. of Physics, University of Durham, Durham, DH1 3LE, U.K.*

In the second part, I deal with the stunning information that we derive from studies of high redshift galaxy clusters. These are fundamental observations because they show that at least some galaxies in clusters evolve over relatively short look-back times. The natural interpretation is that we are witnessing the transformation of Spiral galaxies into E and/or S0 galaxies Although mergers are supported by recent images from the Hubble Space Telescope (HST), there are considerable difficulties in reconciling this scenario with the observations of nearby clusters discussed in Section 1.

In the remaining two parts, I have been conscious to turn my attention towards galaxy formation (as opposed to evolution). Section 3 deals with recent theoretical progress that combines our understanding of the hierarchical evolution of gravitational structure with an approximate numerical description of galaxy formation. Because we are now able to place galaxy formation models within a context of cosmological clustering, it is possible to model the effects of galaxy mergers and other processes that transform galaxy morphology. Although this work is still in its early stages, it bears interesting comparison with observation.

The final part of these notes considers how the environmental modulation of galaxy formation distorts our picture of the underlying gravitational structure. If galaxy formation is strongly influenced by the environment, then we must question how the observed spatial variations in galaxy number relate to the underlying mass density fluctuations. Such 'biassing' of the galaxy density field seems an obvious fact: because of the connection between galaxy density and morphology, if we use different types of galaxy to map the universe we will inevitably obtain different results. Rather than treating this phenomenologically, we should now seek to understand the physics that lies behind this bias.

## 1. Galaxy Morphology — Nature vs Nurture

In Hubble's classification of galaxy morphologies, the principal dividing line comes between the early-type galaxies (ie., E and S0) and the spiral (or late-type) galaxies. The essential distinguishing feature is the smoothness of the galaxies' light distribution; this, in turn, reflects the presence or absence of knots of bright star formation. The basic division thus corresponds to one of on-going star formation rate (the rate of star formation being higher in the later types). The same sequence also reflects an increasing contribution from disc light. At one extreme, the light distributon in E galaxies can be described by a pure spheroid. At the other extreme, the Sd and irregular galaxies are best pictured as a pure, but chaotic, disc of very active star formation.

There is a fundamental question that we must address: when did this distinction occur? At first sight, one would expect that elliptical and S0 galaxies should cover the whole range of ages, from systems that had just recently switched off star formation, to systems that formed very rapidly at some early epoch and had remained quiescent ever since. In this picture (Toomre & Toomre, 1972), the star formation in galaxies might be turned off by collisions and mergers: the violence of these events being sufficient to redistribute the disc stars into a smooth spheroid. Less catastrophic collisions, or interaction with the hot, dense gas that is trapped in galaxy clusters, would deplete a galaxy of its gas reservoir so that star formation would be slowly strangled. Larson et al. (1980) suggested that this might make a plausible formation history for S0 galaxies. The generic scenario is commonly referred to as the 'nurture' origin of galaxy morphology because all galaxies are formed similarly and their present-day morphology is later imprinted on them by their environment.

This scenario does not, however, square up to all the observational evidence (a detailed discussion follows below), and it is necessary to invent an alternative scenario. This is best represented by the models of Arimoto and Yoshii (1987). In their scheme, the formation of the bulges (or spheroids) of galaxies occurs as a single event at an early epoch. In some environments, these bulges are then able to accrete gas discs as a secondary part of the formation process. I will refere to this scenario as the 'nature' origin of galaxy morphology.

The two scenarios that I have outlined are extreme views, and the truth probably lies somewhere in between. It is nevertheless useful to think in terms of these simple opposites when considering the evidence summarised below.

*1.1. The Colour–Magnitude Relation*

The distinction between early and late type galaxy morphologies depends only on the *present-day* star formation rate. Since the colours of early-type galaxies younger than 5, and even 10 Gyr, depend strongly on the time elapsed since the formation of the last stars, an investigation of the colours of E and S0 galaxies should uniquely distinguish between the 'nature' and 'nurture' formation scenarios. For example, the U-V colour of a single age population of stars born 5 Gyr ago changes at a rate of about $0.^m05$ per Gyr. Thus if we were to compare two such galaxies with ages of 4 and 6 Gyr, their difference in colour ($\sim 0.^m1$) would be easily detectable.

The experiment is most reliably performed in galaxy clusters, since there is then a large population of E and S0 galaxies that share a common distance. What is found is a remarkably strong correlation between the colours of early-type galaxies and their absolute magnitudes. Since the global trend can be well accounted for as a correlation between metal abundance and the total galactic mass, the 'nature' scenario immediately establishes itself as the front runner. Although this view has recently been challenged by detailed population synthesis studies based on the variation of the ratios of the H$\beta$, Fe and Mg b spectral lines (Faber et al., 1995), it remains the most natural explanation of the observed correlation. Even then, the small scatter observed in the cores of clusters (Bower et al., 1992) is surprising. For example, the rms scatter in (U-V) colour is less than $0.^m04$ in the Coma cluster (Figure 1). This data means that early-type galaxies have to be old (ie., the bulk of their stars formed more than 10 Gyr ago) so that the sensitivity of their colours to their exact formation epoch is reduced to an acceptable level. Importantly, this constraint is insensitive to the stellar population model used in the calculation. The only alternative to forming these galaxies at early times is to somehow coordinate the epoch at which their star formation terminates.

The implications of the colour-magnitude relation cut even deeper than the galaxy formation epoch, however. The small, but significant, slope of the colour–magnitude relation suggests that the evolution (as distinct from formation) of early-type galaxies cannot merge together galaxies of differing absolute magnitudes. If this were to happen, any pre-existing colour-magnitude relation would be washed out, again violating the small scatter observed. It thus appears that an early-type galaxy must 'know' its final mass before it forms the bulk of its stars: this is the only way in which it can arrange to end up with the correct final colour.

*1.2. Galaxies Outside Clusters*

Although I have treated the above discussion as applying to early-type galaxies generally, it strictly applies only to the galaxies in the central parts of clusters. Nevertheless, these samples must include some galaxies that are traversing the cluster core for the first time, and galaxies that appear in the cluster centre only in projection. Thus the statements made about cluster galaxies also constrain galaxies in the outer parts of clusters.

Line strength indices measure essentially the same information as the colours of galaxies (ie., the mean stellar effective temperature) but are better suited to the study of galaxies whose distance is poorly defined. Studying the properties of galaxies in the outer parts of the Coma cluster, Guzman et al. (1992) find a weak trend for the outermost galaxies to have (apparently) younger stellar populations than their counterparts at the cluster centre. Rose et al. (1994) find a similar result comparing the coadded spectra of samples of high and low richness cluster galaxies. These data provide some evidence to suggest that the properties of early-type galaxies really do depend on environment, but it must be emphasised that the trend is weak.

Moving further away from clusters, Schweizer & Seitzer (1992) have studied early-type galaxies that are isolated or in low mass groups. They find a definite correlation between the deviation of galaxies from the mean colour–magnitude (or line strength–velocity dispersion) relation and the degree of 'fine structure'. Images showing such structure (low surface brightness ripples, 'jets' and isophote distortions) are indicative of a merger event in the galaxies' recent past. Thus this data provides direct evidence for the late formation of a number of E and S0 galaxies. In some cases, such as NGC 7252, there is over-whelming evidence that we are witnessing the merger of two gas-rich galaxies, and that in 2–3 Gyr the remnant will settle down to become *morphologically* indistinguishable from a prototypical E galaxy. The time-scale for the colours of such galaxies to relax onto the colour-magnitude relation is, however, of order 5–10 Gyr (if it will indeed ever happen). Thus, except in Schweizer & Seiter's *bulge*-dominated merger models, the lack of substantial numbers of 'young ellipticals' remains a problem. Seemingly, the best resolution is to accept that galaxy mergers do form ellipticals, but then to hypothesise that such mergers were much more common at high (possibly very high) redshifts. In this way, it is perhaps possible to combine the best of both scenarios.

*1.3. Globular Clusters*

Until recently, globular clusters seemed to present a strong arguement in favour of the nature hypothesis. The specific frequency of globular clusters (ie., the number of globular clusters per unit mass of the parent galaxy) is considerably higher in early-type galaxies than in late-type systems of similar mass (Harris & Racine, 1979). The discrepancy is particularly acute in dominant cluster galaxies. This observation presents a serious problem for the 'nurture' scenario unless additional globular clusters are produced during the morphology-determining merger event. Recent observations of interacting and merging galaxies with HST (eg., Whitmore & Schweizer, 1995) show, however, that star clusters are indeed formed. The remaining question is then whether these newly formed clusters will evolve dynamically to become indistinguishable from the earlier globular cluster population.

This raises the possibility of using the properties of the globular cluster population to date the last major merger event in the formation history of ellipticals (eg., Zepf et al., 1995). If the colours of the globular cluster population show a double peaked structure, then this can be taken as an indication that the galaxy has been formed from two (or more) distinct components. The difficulty with this method is that, since globular clusters are known to be metal poor, it is not possible to unambiguously distinguish between their age and metal abundance. The information contained in the globular cluster formation record is thus open to a wide variety of possible interpretations: for example the double-peaked structure might arise from a complex formation history that occurred at high redshift.

### 1.4. Ultraviolet Colours

Observations of nearby elliptical galaxies by the IUE satellite show that they emit a surprisingly large far ultraviolet flux (Burstein et al., 1988). Furthermore, there are wide variations between galaxies and a suggestion that the flux is correlated with the strength of the $Mg_2$ spectral index. The origin of this ultra-violet continuum has not yet been solved. While one possibility is that it is produced by a small population of young stars, it is also possible that it may be produced by some extreme forms of evolution in older stars (Greggio & Renzini, 1990): unfortunately it is extremely difficult to model the required stellar evolutionary tracks with sufficient accuracy. At present, the young star hypothesis is the less favoured (cf., Dorman et al., 1995) because galaxies that are clearly undergoing bursts of star formation fail to show the expected UV flux. However, if the extreme stellar evolution hypothesis is correct, the wide variations between galaxies in the extent of the UV up-turn is still a mystery.

### 1.5. Colour and Line-Strength Gradients

I refer here to the gradient in these quantities within individual elliptical galaxies. The observed gradients are weak, and this presents a serious challenge for theories that form early-type galaxies in a single burst of star formation. In isolated gas-cloud models, such as that of Larson (1975), star formation is accompanied by the infall of gas towards the centre of the galaxy. As the gas falls towards the centre, its metal abundance rises. Thus the stars formed on the outside of the system tend to be metal poor, where as those at the centre are highly metal rich. These gradients are incompatible with current observations (eg., Peletier et al., 1990, Davies et al., 1993): in order to rescue the model, it is necessary to suppose that the gas is stirred up by supernova explosions or galaxy mergers. Some such process must in any case be at work in order to extract angular momentum from the infalling gas thus preventing it from forming a disc.

Another factor that needs to be considered is the variation in gradient from one galaxy to another. Although this initially presented a challenge to the 'nature' view-point, Franx & Illingworth (1990) argued that most early-type galaxies were consistent with a single relationship between colour (or line strength) and the local escape velocity (ie., that rotation and surface brightness need to be taken into account when comparing galaxies). The detailed variations of different metal lines remain difficult to account for, however (Fisher et al., 1995). Once again the data are inconclusive, it being possible to arrange for both scenarios to produce galaxies with the observed properties.

*1.6. Other Issues*

I have collected below a number of other issues that are important to the 'nature' vs. 'nurture' discussion but for which I have run out of the space need to present an adequate discussion.

Firsty, the discussion above has focused on normal elliptical galaxies and the bulges of S0's. It is unclear to what extent the discs of S0 galaxies obey a colour-magnitude relation. Another exception might be the central cD galaxies of rich clusters. It seems probable that their special location at the focus of cooling flows and dynamical friction have had their own distinctive effects on their evolution.

Secondly, clusters of galaxies do contain significant numbers of spirals. These are usually pictured as late-comers to the cluster environment. Environmental effects on spiral galaxies are hard to quantify through optical measurements, but there is definite evidence for a lack of HI gas haloes in spiral galaxies in the Virgo cluster (Giovanelli & Haynes, 1985). It would seem that spiral galaxies that fall into a cluster are robbed of their gas supply. It is not at all clear, however, that the rate at which this happens in Virgo galaxies is consistent with the numbers of early-type galaxies found in low-density environments.

A third issue is the connection between elliptical galaxies and the bulges of spirals. Spiral bulges rotate more rapidly than giant elliptical galaxies, but it now seems clear that their dynamical properties are very much the same as those of small ellipticals. Since the luminosities of spiral bulges are lower than those of typical E galaxies (or indeed the bulges of S0 galaxies), the latter is the more appropriate comparison to make. The present conscensus is thus that bulges *are* small ellipticals (Davies et al., 1993, but see Balcells & Peletier, 1994, for a discussion of colour differences). While this unification does not solve the nature vs. nurture question — it simply requires that spiral galaxies must have had sufficient time to accrete their gas discs since their last merger — it does raise the likelihood of an environmental modulation of the overall galaxy luminosity function. I will return to this issue in Section 3.

Finally, I have found no space to dicuss the structure and dynamics of E galaxies. Two particular issues would warrant extensive discussion: the 'fundamental plane', and the frequent occurance of counter-rotating cores. As a result of the virial theorem, the values of the structural parameters of elliptical galaxies (usually, the surface brightness, luminosity and stellar velocity dispersion) fill only a two-dimensional space: the 'fundamental plane'. The width of this plane may be used to constrain the stellar populations of elliptical galaxies in a similar way to the colour-magnitude relation (cf., Bender et al., 1992). Counter-rotating cores — where the angular momentum of an elliptical galaxy reverses in the central region — are observed in many nearby galaxies. Although there is no way to date the event from this data, the existence of such reversals is strong evidence for mergers at some time during the galaxies' formation history.

## 2. Distant Clusters — Observing Galaxy Evolution Directly

The Butcher-Oemler effect (Butcher & Oemler, 1978) provides a direct demonstration that the population of galaxies within clusters has changed substantially over relatively modest look-back times. As originally stated, the effect relates to the increased fraction of blue galaxies in high and moderate redshift clusters. However, the definition has subsequently been expanded to also include the appearance of starburst and post-starburst (aka., E+A) galaxies (Dressler & Gunn, 1983, Couch & Sharples, 1987). It is tempting to suggest that we are witnessing the epoch at which E galaxies are formed from spirals, or at which star formation in spirals has just begun to fade, converting them into S0 types. Unfortunately, this scenario has a number of problems. For example, the anomalous spectra are not consistent with the passive fading of a spiral galaxy and require the star formation to terminate in a strong burst. Such unusual galaxies are not seen in the cores of rich clusters at the present epoch. The Butcher-Oemler puzzle might therefore be rephrased "why do the blue starburst galaxies disappear from rich clusters by the current epoch?" A full understanding of the Butcher-Oemler effect will require a more complex explanation than the fading of spiral galaxies in the cluster environment.

Following along the same line of argument, there have been several studies of the spatial and velocity distributions of the anomalous galaxies. Although the initial results suggested that the starburst and E+A galaxies were less concentrated to the cluster centre than their normal counterparts (as would be expected if these were galaxies that had only recently fallen into the cluster), more detailed studies have not supported this claim (cf., Allington-Smith et al., 1993).

Finally, the fraction of anomalous galaxies in the high-redshift clusters is uncomfortably high given that their unusual spectra and blue colours are short lived ($\sim 1\,\mathrm{Gyr}$). If the rate of their creation is approximately constant, then 50% of the present-epoch early-type galaxy population would have been produced since a redshift of 0.4. Such a high fraction of 'young' early-type galaxies is inconsistent with the existence of a well-defined colour–magnitude relation. It is only possible to explain the observed homogeneity if the star formation that we observe in these high redshift galaxies is significantly skewed towards high mass stars (there is, however, some evidence for this in local starburst events, Reike et al., 1993). Another possibility is that the starburst phenomenon in these galaxies contributes so much of their luminosity that they subsequently fade below the point at which galaxies are normally included in colour–magnitude studies.

Recently, imaging by the Hubble Space Telescope (eg., Dressler et al., 1994, Couch et al., 1994) has added new spice to the Butcher-Oemler debate. It is now possible to determine accurate morphologies for the anomalous Butcher-Oemler galaxies. The results suggest that interactions between spiral galaxies account for the most active of the starburst galaxies. While, many of the other blue galaxies appear to be normal spirals, the red galaxies with anomalous 'E+A' spectra resemble present-day E galaxies. Since the duration of strong tidal features and ripples (the dynamical evidence of a recent merger) is probably shorter than the lifetime of the A stars formed during the starburst (Mihos, 1995), the HST data would seem to lend support to the fading interpretation. Deeper images (with WFPC2) are needed in order to set better limits on the tidal disturbance of the E+A galaxies.

It is not at all clear, however, why the environment in these high redshift clusters is so conducive to galaxy mergers. This has prompted a more thorough investigation of the outer parts of nearby clusters. Indeed, Caldwell et al. (1993) discovered a population of E+A galaxies apparently associated with the NGC 4839 group that is being tidally disrupted as it falls into the Coma cluster. We must beware, however, of moving goal posts: the spectra of these low redshift Butcher-Oemler galaxies are certainly not as spectacular as those seen in the distant clusters.

Before leaving this subject, one final comment is needed. The extensive discussion of the Butcher-Oemler effect galaxies often leaves the impression that all galaxies in distant clusters are anomalous. In fact, the opposite is true: Aragon-Salamanca et al. (1993), showed that even clusters at $z = 0.5$ have a well-defined colour–magnitude relation.

The constraints based on the color-magnitude relation of nearby clusters become even more imposing in these high redshift systems. Stanford et al. (1995) have determined the optical–infrared colour–magnitude relation in two distant clusters, A370 at $z = 0.374$ and A851 at $z = 0.407$. Despite the large look-back time, the scatter they measure is less than 0.07 mag for E/S0 galaxies selected via Hubble Space Telescope morphologies. Ellis (1995) reports similar results for rest-frame U-V colours in the 0016+16 cluster at $z = 0.55$. Reworking the analysis of Bower et al. (1992), this pushes the formation epoch for these ellipticals back to beyond 12 Gyr ago. In addition, the colour-magnitude relation at these redshifts has the same form as that observed locally. While there is a blueward shift (as expected from passive stellar evolution), the slope of the relation remains close to the present-day value. This reinforces the view that the slope of the relation is driven by metal abundance differences.

## 3. Understanding the Morphology–Density Relation

So far I have concentrated on observation, now some recent theoretical progress deserves discussion. The development of analytical models for the evolution of gravitationally bound structures has opened up the possibility of developing quantitative models of the morphology–density relation. The background gravitational evolution, which describes how galaxies are built up from sub-galactic units and then incorporated into larger and larger groups and clusters, is provided by extensions of the Press-Schechter theory (Press & Schechter, 1974, Bower, 1991, Bond et al., 1991). This allows a large number of Monte-Carlo histories (Figure 2) to be generated rapidly, allowing the huge parameter-space of galaxy formation models to be efficiently explored. For example, with a given set of assumptions about the efficiency of galaxy mergers, Kauffmann & White (1993) were able to rapidly calculate the dependence of galaxy morphology on environment by varying the background density that is used to generate the halo merger history. Nevertheless, unless the prescription for galaxy formation and the influence of environment is kept simple, too many unconstrainable parameters are introduced. The challenge is thus to develop a simple model that is able to reproduce both the abundances of the various galaxy types (and their modulation by the environment) and the considerable homogeneity of the colour–magnitude relation of cluster E galaxies.

One key ingredient for these models is the feed-back of kinetic energy from supernovae into the gas reservoir surrounding the galaxy (Matthews & Baker, 1971). This heating of the gas counterbalances its cooling and thus regulates the rate at which stars can form. Without this mechanism, all the available baryonic material would rapidly be locked into stars almost immediately after the universe recombines. However, with a single regulatory parameter, (set to produce galaxies with the correct light-to-mass ratio) too many faint galaxies are generically produced (Blanchard et al., 1992, Kauffmann et al., 1993). Possibly, the effects of a photoionizing background needs to be included — this would suppress the formation of the dwarf galaxies because of their low optical depth — or else, some of the dwarf galaxies may become unbound when their gas is expelled by the supernova driven wind.

Another outstanding problem for these models is the need to establish a more amenable definition of galaxy morphology. Hubble's distinction between the galaxy types involves many subjective factors, such as the predominance of spiral arms. In order to allow comparison with theoretical models, a better quantified definition is required: for example, one based on the bulge-to-disk light ratio in the B photometric band.

Extending these models to trace the evolution of metal abundance adds an important extra constraint. In the work of Arimoto & Yoshii (1987), elliptical galaxies were assumed to form their stars in a single burst, and thus the depth of their gravitational potential, plus the feedback from supernovae, was able to create a well defined colour–magnitude relation that approximately matched that observed. This scenario differs considerably, however, from the continuous star formation envisaged in the models considered above. Kauffmann (1995) shows that the biased formation of massive galaxies produce a correlation between metal abundance and galaxy magnitude, but it is not yet clear whether the relationship is tight enough to match with the observed colour-magnitude relation. This is a key question for future work.

## 4. How Biased is Galaxy Formation?

Since galaxy morphology depends on environment, galaxies (even if all types are combined) cannot be relied upon to trace the mass distribution of the universe. This seems obvious, yet the galaxy luminosity function is to a remarkably good approximation universal (eg., Loveday et al., 1995). The paradox is a major puzzle for modern cosmology. If early-type galaxies had only recently gained their distinct morphology, the discrepancy would be minimised. However, we have seen that there are wide differences in the stellar populations of early- and late-type galaxies. This affects their luminosities both through their mean colours and through the length of time over which their stars have been able to form. The problem is worse, however, if elliptical galaxies and the bulges of spirals form early in the universe. In this scenario, only galaxies outside clusters are able to accrete gas discs, and thus their luminosity continues to increase upto the present day. In order for such a scenario to be consistent with a universal luminosity function, a cosmic conspiracy is required: the smaller initial luminosities of field spiral bulges need to be counterbalanced by the total number of stars formed in their disks.

Considering these possibilities brings us to wonder if we really know the true range of galaxy properties (cf., Philips et al., 1987, McGaugh et al., 1995). Perhaps the perceived universality of the luminosity function is in part due to our preconceived ideas about the appearance of galaxies in low density regions. Much recent work has concentrated upon searches for low surface brightness galaxies. Some have indeed been found, but not in prodigious numbers. Furthermore, they appear to follow the spatial distribution of conventional galaxies (Eder et al., 1989, Mo et al., 1994): they do not contribute a population of 'failed' galaxies that fill out the holes in the large scale galaxy distribution.

The voids in the galaxy distribution raise the possibility that the effects of feedback from galaxy formation might propagate over significant cosmological distances. For example, the inter-galactic ultra-violet radiation field might have played an important role in suppressing the formation of small galaxies (eg., Efstathiou, 1992). If we stretch this argument to its extreme, we might question whether the galaxy distribution is only tenuously connected with underlying mass distribution (Babul & White, 1991, Bower et al., 1993). Fortunately, it appears unnecessary appeal to such phenomena in order to understand observational data on large-scale structure: a 'global' bias that depends only on the local density is sufficient. However, this phenomenological approach still begs the question "what happened to the galaxies that failed to form in the voids?"

## Summary

**Nearby galaxies.** The colour–magnitude relation observed for early-type galaxies sets strong limits on the numbers of 'young' ellipticals and S0s in clusters. Furthermore, it also rules out substantial growth by mergers after the majority of stars have been formed. Against this, 'young' early-type galaxies are present in the field. Is it possible that there are two sorts of elliptical galaxy? If so, it seems reasonable to hope that we can distiguish the early- and late-forming galaxies in some way other than through their colours. One interesting possibility is that the globular cluster populations of E galaxies might provide a merger history, each peak in the colour histogram corresponding to a separate galaxy-merger event. On balance, however, the majority of stars in elliptical galaxies and the bulges of spirals would seem to be old. Currently, the most appealing scenario is one in which star formation activity has declined since some early initial epoch of galaxy formation. If this rate of decline were much steeper in the dense (proto-) cluster environment, the homogeneity of early-type galaxies in clusters could be explained, without contradicting the discovery of small numbers of 'young' ellipticals in the field. Galaxy mergers might also have been prevalent during the early initial burst of star formation, and this could account for the low angular momentum and small colour gradients of bulges without contradicting the correspondence between total mass and galaxy colour.

**The Butcher-Oemler effect.** The major puzzle surrounding the Butcher-Oemler effect is not so much to suggest what might cause the starburst phenomenon, as to explain why it goes away by the present epoch. HST imaging suggests that mergers play an important role in promoting the violent star formation activity, but it is unclear why the cluster environment at $z = 0.5$ is so conducive to galaxy mergers. Recent observations have suggested that there is a weak counterpart to this effect in groups of galaxies falling into clusters at the present-day. Overall, the effect lies in stark contrast to the apparent uniformity of present-day cluster galaxies. In order to reconcile the observations, it is necesary to suggest that the newly formed stars have an initial mass function strongly skewed to high masses, or that the starburst activity occurs in systems of low intrinsic surface brightness.

**Theoretical progress.** The observational existence of a galaxy morphology–density relation and and a colour–magnitude sequence have been known for over two decades. At last, our theoretical understanding is catching up. Progress in modelling the hierarchical growth of gravitational structure is now being applied to explore a wide variety of galaxy formation and evolution scenarios. There are still extensive problems to be solved — for example, a better quantified definition of galaxy morphology needs to be established, and the stellar evolution codes need to incorporate the development of the galactic metal abundance — but it is becoming possible to make detailed quantitative tests of galaxy formation schemes.

**Biased galaxy formation.** Aside from its own intrinsic interest, an understanding of the density–morphology relation is required before we can reliably interpret maps of the universe made by observing galaxies. In particular, the two-stage galaxy formation models that seem best able to account for the properties of nearby galaxies jar against the perceived universality of the overall galaxy luminosity function. On a positive note, if we understood the origin of the density-morphology relation, we could select particular types of easily observed galaxies with which to map the universe, using our knowledge to undo this deliberate bias at a later stage.

Drawing these strands together, it appears that the data are forcing us to develop a hybrid model in which mergers do play an important role in the morphology of galaxies, but in which the majority of stars (in cluster galaxies at least) are formed at early epochs. The theoretical developments are very encouraging in this respect. They now need to be pushed harder to make more detailed comparisons with the observable properties of galaxies. The models also await to be combined with studies of large-scale structure. In this way, it will be possible to make a quantitative test of 'biased galaxy formation'.The Butcher-Oemler effect still remains a puzzle in all this, but the theoretical models can help us here too by making predictions about the way in which the effect should develop at even higher redshifts and how it should be modulated in other environments.

My brief for writing these notes was to address the influence of the environment on galaxy formation. There remain two competing theories, both of which come close to explaining the data. In the first, pure disk galaxies are the basic building blocks of the luminous universe. In regions of high density, but low relative velocities, collisions rapidly create the spheroidal components that then come to dominate the cluster environment. The second model is subtly, but radically different, most stars in the universe are formed at an early epoch, when mergers of galaxy sized haloes are common. Only in low density environments are these then able to accrete gas disks. The roles of spheroids and disks are thus reversed in these two models: it is a challenge to the workshop to decide which came first, the spheroid or the disk.

### Acknowledgements

In this section, I would like to acknowledge all the authors whose papers constitute the background of these notes. In order to keep the reference list to reasonable the length, I have had to restrict my citations to papers that are either seminal, or that have contributed a recent major advance in the subjects. In many cases, the choice has probably been made unfairly. I would also like to thank the numerous people that have shared their ideas with me. I am grateful to acknowledge the use of the Babbage electronic preprint server at SISSA, and the BIDS electronic citation index. Finally, I would like to thank the conference organisers for giving me the opportunity to collect my thoughts on these subjects together into a coherent whole.

## Figure Captions

**Figure 1.** The colour–magnitude relation for early-type galaxies in the Coma cluster. Data are taken from Bower et al. (1992). All but two elliptical galaxies sit on a tightly defined correlation between U-V colour and total V magnitude. The scatter among the S0 galaxies is slightly larger, but the relation is still well defined.

**Figure 2(a–d).** Monte-Carlo trajectories showing the formation of dark-matter haloes. The final masses of the haloes are, respectively, 0.1, 1.0, 10.0 and 100.0 times the mass of a typical halo at the final epoch. The plots correspond approximately to the halo masses of an isolated galaxy, a small galaxy group, a large group and a cluster. It can be seen that low final-mass objects form early in the history of the universe, and then are left behind by the hierarchical growth of other structures. On the other hand, high final-mass objects are forming very rapidly at the current epoch.

Time is given in units of the final age of the universe, and mass is given as a fraction of the characteristic halo mass at the final epoch. The calculations use the method described by Lacey & Cole (1994) tracing the larger fragment at each subdivision. The universe is assumed to have critical density, and the mass fluctuations are characterised by a power-law spectrum with index $-1$.

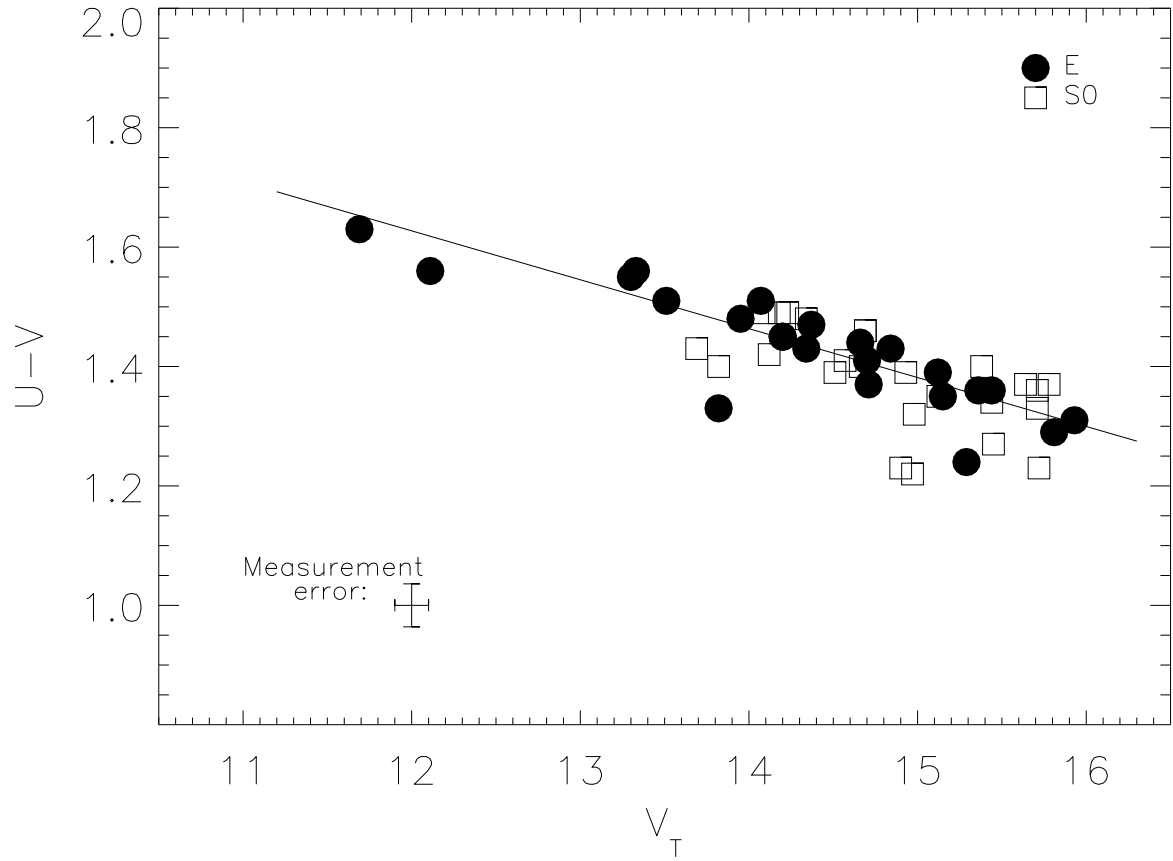

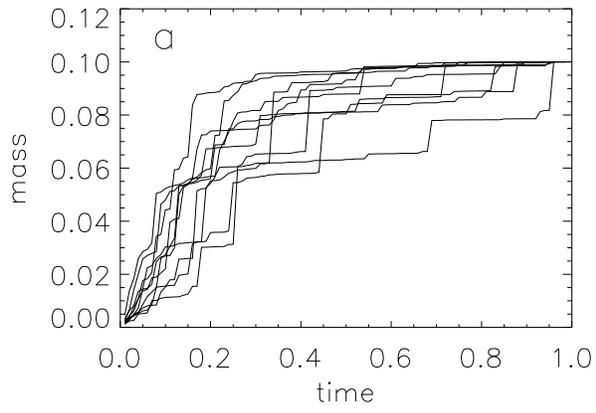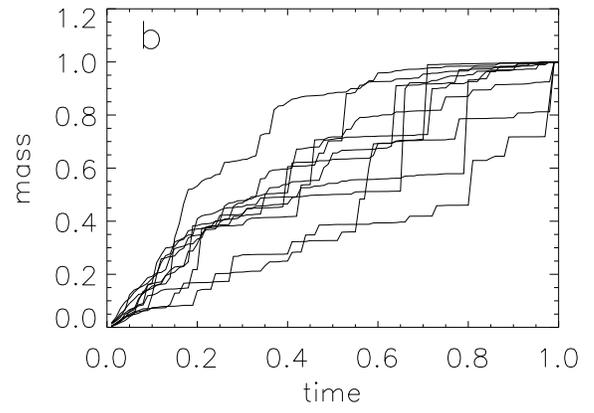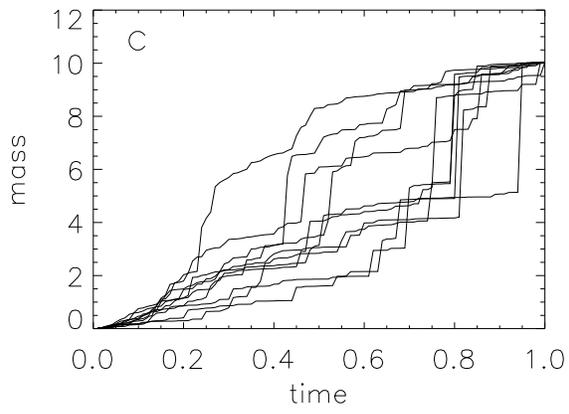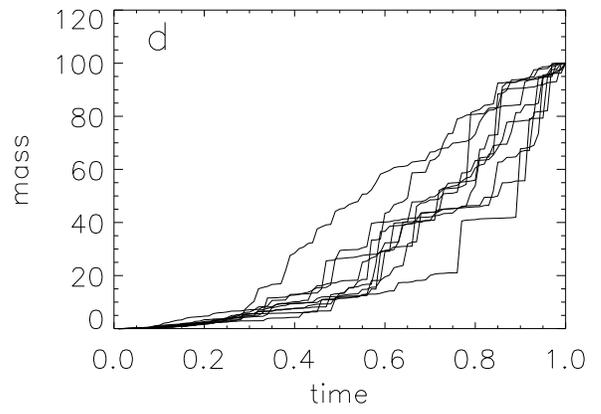